# An effective parameterization of texture-induced viscous anisotropy in orthotropic materials with application for modeling geodynamical flows

Javier Signorelli[1], Riad Hassani[2], Andréa Tommasi[3], and Lucan Mameri[3]

[1] Instituto de Física de Rosario, CONICET & Universidad Nacional de Rosario, Argentina
[2] Université Côte d'Azur, CNRS, Observatoire de la Côte d'Azur, IRD, Géoazur, France
[3] Géosciences Montpellier – CNRS & Université de Montpellier, France

We describe the mathematical formulation and the numerical implementation of an effective parameterization of the viscous anisotropy of orthorhombic materials produced by crystallographic preferred orientations (CPO or texture), which can be integrated into 3D geodynamic and materials science codes. Here, the approach is applied to characterize the texture-induced viscous anisotropy of olivine polycrystals, the main constituent of the Earth's upper mantle. The parameterization is based on the Hill (1948) orthotropic yield criterion. The coefficients of the Hill yield surface are calibrated based on numerical tests performed using the second order Viscoplastic Self-consistent (SO-VPSC) model. The parameterization was implemented in a 3D thermo-mechanical finite-element code developed to model large-scale geodynamical flows, in the form of a Maxwell rheology combining isotropic elastic and anisotropic non-linear viscous behaviors. The implementation was validated by comparison with results of the analytical solution and of the SO-VPSC model for simple shear and axial compression of a homogeneous anisotropic material. An application designed to examine the effect of texture-induced viscous anisotropy on the reactivation of mantle shear zones in continental plates highlights unexpected couplings between localized deformation controlled by variations in the orientation and intensity of the olivine texture in the mantle and the mechanical behavior of the elasto-viscoplastic overlying crust. Importantly, the computational time only increases by a factor 2-3 with respect to the classic isotropic Maxwell viscoelastic rheology.

## 1 Introduction

The greatest challenge in modeling the dynamics of the solid Earth is to reproduce the extremely heterogeneous deformation of the outer layer of the planet – the tectonic plates, which have thicknesses ranging from a few km at oceanic ridges to > 200 km beneath some continental domains, in response to the continuous motion that animates the underlying convective mantle. An additional challenge is to simulate the repeated reactivation of some plate boundaries with intervals of hundreds of million years, (*e.g.*, Wilson 1966; Vauchez et al. 1997). This behavior implies long-time preservation of rheological heterogeneity produced by deformation in response to episodes of collision or breakup of tectonic plates. Two processes have been proposed as the causes for this behavior: (1) grain size reduction (Bercovici and Ricard 2014) and (2) viscous anisotropy (Tommasi and Vauchez 2001; Tommasi et al. 2009). Here, we focus on the second process. All major rock-forming minerals have low symmetries (hexagonal, trigonal, orthorhombic, monoclinic, or triclinic) and, by consequence, strongly anisotropic thermomechanical behaviors. In addition, tectonic plates deform mostly by non-linear viscoplastic processes, mainly dislocation creep. Deformation produces therefore strong preferred orientations of the crystals that compose the rocks (texture), which transfer part of the intrinsic anisotropy of the physical properties of the rock-forming minerals to larger scales (texture-induced anisotropy).

Olivine, which is the major constituent (60 − 80 %) of the, in most cases, strongest section of the tectonic plates (50 − 90 %of the total thickness of the plate), is orthorhombic. It displays a marked anisotropy of its thermal and mechanical properties (Tommasi et al. 2001; Abramson et al. 1997; Bai et al. 1991). Geodynamical flows produce olivine textures, which are only significantly modified (strengthened or weakened and have their orientation and symmetry changed) by further deformation, (*e.g.*, Nicolas and Christensen 1987; Wenk et al. 1991; Tommasi et al. 2000). Seismological measurements are able to detect the elastic anisotropy produced by these textures, (*e.g.*, Hess 1964; Savage 1999; Tommasi and Vauchez 2015). These data record patterns of active or old (fossilized) olivine textures, which are homogeneous at the scale of several tens to hundreds of km in the upper mantle. These textures produce viscoplastic anisotropy at large-scale (Knoll et al. 2009; Hansen et al. 2012; Mameri et al. 2019), which in turn modifies subsequent deformation. Texture-induced viscous anisotropy has been proposed as a major feature of plate tectonics, explaining the reactivation of ancient structures, even at hundreds of million years of interval (Vauchez et al. 1997; Tommasi and Vauchez 2001; Tommasi et al. 2009). It has also been proposed to modify the convective patterns and the interactions between the plates and the convective mantle (Castelnau et al. 2009; Lev and Hager 2011; Blackman et al. 2017). A precise description of the viscous anisotropy produced by olivine textures was achieved by explicitly coupling viscoplastic self-consistent calculations of polycrystal deformation into geodynamical flow models (Knoll et al. 2009; Tommasi et al. 2009; Castelnau et al. 2009; Blackman et al. 2017). However, this approach remains too time and memory-consuming for widespread use.

Hierarchical multi-scale modelling provides a path for exchanging information between systems with different characteristic length scales. Several strategies may be defined (Gawad et al. 2015; Jiang 2014). In the present work, we adopt the virtual-experiment strategy for exploring the viscosity tensor of a textured material composed of orthorhombic crystals, using crystal plasticity modeling (i.e. a micromechanical crystal plasticity model captures texture-induced evolution of plastic anisotropy). Since simulations are able to calculate the material response at any point of the yield surface, the virtual-experiment strategy provides a method for identifying the parameters in the yield function for deformation modes. This strategy allows combining the accuracy of the description of the texture-induced viscous anisotropy of the polycrystal model and the numerical performance of the yield functions (Plunkett et al. 2006).

In this article, we describe the mathematical formulation and the numerical implementation of a parameterization of the anisotropic viscoplastic rheology of olivine polycrystals in a 3D finite-element thermo-mechanical code (Adeli3D, Hassani et al. 1997). The parameterization is based on the Hill (1948) orthotropic yield criterion in which the coefficients that calibrated the Hill yield surface are based on numerical tests performed using the second order Viscoplastic Self-Consistent (SO-VPSC) model. The aim is to provide a simplified Maxwell rheology model, combining isotropic elastic and anisotropic non-linear viscous behavior, that can be integrated into 3D geodynamic and materials science codes, allowing the physical description of a complex multi-scale system with reasonable computational times.

## 2 Parameterizing texture-induced viscous anisotropy in viscoelastic orthotropic materials

The framework presented in this section to parameterize the anisotropic rheology of an olivine polycrystal is based on the notion of hierarchical multi-scale modelling. It extends the classical viscoelastic Maxwell model to consider viscous anisotropy calibrated based on the Hill (1948) orthotropic yield criterion. The parameters describing the anisotropy at the coarse-scale (plate tectonics) are identified using fine-scale (rock sample) virtual experiments performed using viscoplastic self-consistent simulations of the deformation of an olivine polycrystal.

### 2.1 Thermo-mechanical code

We implemented the viscous anisotropy parameterization in the 3D thermo-mechanical code Adeli3D (Hassani et al. 1997), which is optimized to model large-scale geodynamical flows, focusing on the deformation of the lithosphere, i.e. the semi-rigid plates that compose the outer layer of the Earth. The code is based on a Lagrangian finite-element discretization of the quasi-static mechanical behavior of the lithosphere, which solves the obtained non-linear equations using a dynamic relaxation method (Cundall 1988). The problem consists in finding the velocity field **v** and the symmetric tensor $\sigma$ satisfying

$$\begin{cases} \mathbf{div}\,\sigma + \rho_\ell \mathbf{g} = \mathbf{0} \\ \dfrac{D\sigma}{Dt} = \mathcal{M}(\sigma, \mathbf{D}), \end{cases} \quad \text{in } \Omega \qquad (1)$$

in the physical domain $\Omega$, where **g** is the acceleration vector due to gravity, $\rho_\ell$ is the lithosphere density and $2\mathbf{D} = \nabla \mathbf{v} + \nabla \mathbf{v}^\top$, $\dfrac{D}{Dt}$ is an objective time derivative associated with the large strain formulation (Malvern 1969). Jaumann and Green-Naghdi are the most frequently used objective derivatives (see, for example, (Johnson and Bammann 1984)). They can be expressed using the same formulation:

$$\dfrac{D\sigma}{Dt} = \dot{\sigma} - \mathbf{W}\sigma + \sigma\mathbf{W}, \qquad (2)$$

with $\mathbf{W} = \omega$, for the Jaumann derivative and $\mathbf{W} = \Omega$, for the Green-Naghdi derivative, where

$$\omega_{ij} = \dfrac{1}{2}\left(\dfrac{\partial V_i}{\partial x_j} - \dfrac{\partial V_j}{\partial x_i}\right) \qquad (3)$$

is the material spin tensor, i.e. the screw part of the velocity gradient $\dfrac{\partial V_i}{\partial x_j}$, and $\Omega = \dot{\mathbf{R}}\mathbf{R}^\top$, the rotation rate associated with the left polar decomposition of the deformation gradient **F** (**F** = **RU**, where **R** is an orthogonal tensor and **U** is a symmetric tensor called right stretch tensor). As generally done, we use the Jaumann corotational stress rate in our simulations.

The functional $\mathcal{M}$ in Equation (1) stands for a general hypoelastic[1] constitutive law. In the present case, $\mathcal{M}$ describes a Maxwell rheology combining isotropic elastic and anisotropic non-linear viscous behavior. The transition between the elastic and viscous regimes depends on the temperature and stress, which change the viscosity, and on the considered time-scale.

### 2.2 Maxwell viscoelasticity

The viscoelastic constitutive law combines the contributions of the elastic $\mathbf{D}_e$ and the viscous $\mathbf{D}_v$ strain-rates. The total strain-rate is defined as $\mathbf{D} = \mathbf{D}_e + \mathbf{D}_v$ and the constitutive relationship is given by

$$\dfrac{D\sigma}{Dt} = 2\mu(\mathbf{D} - \mathbf{D}_v) + \lambda \operatorname{tr}(\mathbf{D} - \mathbf{D}_v)\mathbf{I} \qquad (4)$$

with **I**, the second-order identity tensor, tr, the trace operator, and $\mu$ and $\lambda$, the Lamé moduli. The viscous strain-rate $\mathbf{D}_v$ is given by the flow rule

$$\mathbf{D}_v = \dfrac{\partial \Phi}{\partial \sigma} \qquad (5)$$

where $\Phi(\sigma)$ is a viscous potential. A typical choice for this potential is the power-law equation

$$\Phi(\sigma) = \dfrac{2}{3}\dfrac{\gamma}{n+1}J(\sigma)^{n+1} \qquad (6)$$

where $J(\sigma)$ is the equivalent stress, defined based on Von Mises yield function.[2] In the isotropic case, $\gamma = \gamma_0 \exp(-Q/RT)$, where $n$, $Q$ and $\gamma_0$ are the experimentally-derived power-law exponent, activation energy in $\text{J}\,\text{mol}^{-1}\,\text{K}^{-1}$, and fluidity[3] in $\text{Pa}^{-n}\,\text{s}^{-1}$, respectively; $R$ is the gas constant and $T$, the temperature in Kelvin. Using Equation (5) and Equation (6), the viscous part of the total strain-rate can be expressed as

$$\mathbf{D}_v = \frac{2}{3}\gamma J(\boldsymbol{\sigma})^n \frac{\partial J(\boldsymbol{\sigma})}{\partial \boldsymbol{\sigma}}. \tag{7}$$

---

[1] In hypoelastic constitutive models, the objective stress rate is related to the rate of deformation through an elasticity tensor.

[2] The Von Mises yield function defines a five-dimensional (for an incompressible material) surface in the six-dimensional space of stress: $2J^2(\sigma) = (\sigma_{11} - \sigma_{22})^2 + (\sigma_{22} - \sigma_{33})^2 + (\sigma_{33} - \sigma_{11})^2 + 6(\sigma_{12}^2 + \sigma_{23}^2 + \sigma_{31}^2)$, $J(\sigma) > 0$. When the stress state lies on this surface the material is said to have reached the yield condition.

[3] The fluidity is a material parameter, which depends on the physico-chemical conditions of the deformation, relating the deviatoric stress to the strain-rate.

To describe an anisotropic viscous behavior, the yield condition is here defined based on the orthotropic yield function of Hill (1948).

### 2.2.1 Hill (1948) orthotropic yield criterion

The Hill yield function, which builds on the Von Mises' concept of plastic potential, defines a pressure-independent homogeneous quadratic criterion. It assumes that the reference axes are the principal axes of anisotropy of the material, which are orthogonal. The resulting yield function takes the form

$$J(\boldsymbol{\sigma}) = \left(F(\sigma_{11} - \sigma_{22})^2 + G(\sigma_{22} - \sigma_{33})^2 + H(\sigma_{33} - \sigma_{11})^2 + 2L\sigma_{12}^2 + 2M\sigma_{23}^2 + 2N\sigma_{31}^2\right)^{1/2}. \tag{8}$$

To predict the viscous contribution of anisotropic materials under multiaxial stress conditions, $J(\boldsymbol{\sigma})$ is considered as the definition of the equivalent stress and Equation (7) becomes

$$\mathbf{D}_v = \gamma J(\boldsymbol{\sigma})^{n-1} \mathbf{A} : \mathbf{S} \tag{9}$$

where $\mathbf{S}$ is the deviatoric part of $\boldsymbol{\sigma}$ and $\mathbf{A}$ is a rank-4 tensor describing the material anisotropy, and the stress dependency of $J$ is omitted in order to simplify the notation. Using the Voigt notation, $\mathbf{A}$ has the following matrix representation in the reference frame defined by the principal axes of anisotropy of the material:

$$\mathbf{A} = \frac{2}{3}\begin{bmatrix} F+H & -F & -H & 0 & 0 & 0 \\ -F & G+F & -G & 0 & 0 & 0 \\ -H & -G & H+G & 0 & 0 & 0 \\ 0 & 0 & 0 & L & 0 & 0 \\ 0 & 0 & 0 & 0 & M & 0 \\ 0 & 0 & 0 & 0 & 0 & N \end{bmatrix}, \tag{10}$$

where $F$, $G$, $H$, $L$, $M$, and $N$ are the Hill yield surface coefficients, which describe the anisotropy of the material. In the case of an isotropic aggregate, $F = G = H = 1/2$ and $L = M = N = 3/2$, and Equation (8) reduces to the Von Mises yield function. With this particular choice of the potential, the viscous strain-rate is traceless and the Maxwell constitutive law takes the form

$$\frac{D\boldsymbol{\sigma}}{Dt} = 2\mu \mathbf{D} + \lambda \operatorname{tr}(\mathbf{D})\mathbf{I} - 2\mu\gamma J^{n-1}\mathbf{A} : \mathbf{S}. \tag{11}$$

### 2.2.2 Numerical integration

It is convenient to split the constitutive law given in Equation (4) in deviatoric and hydrostatic parts that can be integrated separately:

$$\dot{\mathbf{S}} + \mathbf{j}_\omega(\mathbf{S}) = 2\mu(\mathbf{D}' - \mathbf{D}'_v) \tag{12a}$$
$$\dot{p} = K \operatorname{tr}(\mathbf{D}) \tag{12b}$$

where $\mathbf{D}'$ and $\mathbf{D}'_v$ ($= \mathbf{D}_v$) are the deviatoric parts of the total and viscous strain-rate tensors, respectively, $p$ is the mean pressure, and $\mathbf{j}_\omega(\mathbf{S}) = \mathbf{S}\boldsymbol{\omega} + (\mathbf{S}\boldsymbol{\omega})^\top$ when the Jaumann derivative is used; $\mu$ and $K$ are the shear and bulk moduli of the material, respectively.

Assuming $\mathbf{D}$ and $\boldsymbol{\omega}$ are constant in the time interval $[t, t+\Delta t]$ and using the Crank-Nicolson scheme, the stress at time station $t + \Delta t$ is sought as the solution of the non-linear equation:

$$\mathbf{S}^{t+\Delta t} = \mathbf{S}^t + 2\mu\Delta t \mathbf{D}' - \mu\Delta t \mathbf{D}'_v(\mathbf{S}^t) - \frac{\Delta t}{2}\mathbf{j}_\omega(\mathbf{S}^t) - \mu\Delta t \mathbf{D}'_v(\mathbf{S}^{t+\Delta t}) - \frac{\Delta t}{2}\mathbf{j}_\omega(\mathbf{S}^{t+\Delta t}), \tag{13a}$$
$$p^{t+\Delta t} = p^t + K \operatorname{tr}(\mathbf{D}). \tag{13b}$$

Equation (13a) can be expressed as $\mathbf{r}(\mathbf{S}^{t+\Delta t}) = \mathbf{0}$, with

$$\mathbf{r}(\mathbf{S}) = \mathbf{S} + \mu \Delta t \mathbf{D}'_v(\mathbf{S}) + \frac{\Delta t}{2} \mathbf{j}_\omega(\mathbf{S}) - \tilde{\mathbf{S}} \tag{14}$$

where $\tilde{\mathbf{S}}$ groups together the terms known at the beginning of the time interval ($t$):

$$\tilde{\mathbf{S}} = \mathbf{S}^t + 2\mu \Delta t \mathbf{D}' - \mu \Delta t \mathbf{D}'_v(\mathbf{S}^t) - \frac{\Delta t}{2} \mathbf{j}_\omega(\mathbf{S}^t). \tag{15}$$

Defining $\mathcal{R}$ as the coordinate system fixed to the laboratory frame and $\hat{\mathcal{R}}$, the corresponding axes fixed related to the material anisotropy (material coordinate system) in this case, a second-order tensor $\hat{\mathbf{X}}$ defined in $\hat{\mathcal{R}}$ can be expressed in $\mathcal{R}$ through a rotation matrix $R_0$, which allows changing from the material to laboratory reference frame (i.e., from $\hat{\mathcal{R}}$ to $\mathcal{R}$):

$$\hat{\mathbf{X}} = \mathbf{R}_0^\top \mathbf{X} \mathbf{R}_0. \tag{16}$$

Expressing Equation (14) and Equation (15) in the material reference frame $\hat{\mathcal{R}}$, we obtain

$$\hat{\mathbf{r}}(\hat{\mathbf{S}}) = \hat{\mathbf{S}} + \mu \Delta t \hat{\mathbf{D}}'_v(\hat{\mathbf{S}}) + \frac{\Delta t}{2} \mathbf{j}_{\hat{\omega}}(\hat{\mathbf{S}}) - \hat{\tilde{\mathbf{S}}} \tag{17}$$

$$\hat{\tilde{\mathbf{S}}} = \hat{\mathbf{S}}^t + 2\mu \Delta t \hat{\mathbf{D}}' - \mu \Delta t \hat{\mathbf{D}}'_v(\hat{\mathbf{S}}^t) - \frac{\Delta t}{2} \mathbf{j}_\omega(\hat{\mathbf{S}}^t) \tag{18}$$

where $\hat{\mathbf{S}} = \mathbf{R}_0^\top \mathbf{S} \mathbf{R}_0$ and $\hat{\mathbf{D}}' = \mathbf{R}_0^\top \mathbf{D}' \mathbf{R}_0$. Newton's method is used to solve the non-linear system $\hat{\mathbf{r}}(\hat{\mathbf{S}}) = \mathbf{0}$ and the inverse rotation is applied to find the deviatoric stress at $t + \Delta t$ in the laboratory reference frame.

The numerical implementation in Adeli3D (hereinafter referred as Adeli3D-anis) of the procedure described in this section allows a parameterized description of the viscous anisotropy component of any orthotropic viscoelastic material. The implementation was validated by comparing numerical and semi-analytical solutions for simple settings (see Appendix A).

## 2.3 Determining the Hill coefficients for olivine polycrystals using the VPSC model

The constitutive relationship presented above includes the effects of anisotropy through the Hill coefficients. In Materials Sciences, these coefficients are determined through a set of mechanical tests. However, such tests are unfeasible for geological materials because viscous deformation is only attained experimentally at high confining pressures, limiting the possible geometry of laboratory experiments. Thus, we calibrate the Hill coefficients, which describe the texture-induced viscous anisotropy of olivine polycrystals, based on VPSC polycrystal plasticity simulations.

Unlike the upper and lower bound models, the VPSC formulation allows each grain to deform according to its orientation and the strength of the interaction with its surroundings. Each grain is considered as an ellipsoidal inclusion surrounded by a homogeneous effective medium that has the average properties of the polycrystal. Several choices are possible for the linearized behavior at grain level. The secant (Hill 1965; Hutchinson 1976), affine (Ponte Castañeda 1996; Masson et al. 2000), and tangent approaches (Molinari et al. 1987; Lebensohn and Tomé 1993) are first-order approximations, which disregard higher-order statistical information inside the grains. However, for highly anisotropic materials displaying a strong contrast in mechanical behavior between differently oriented grains, such as olivine, the second order approximation (SO-VPSC), which takes into account average field fluctuations inside the grains (Ponte Castañeda 2002), is a better choice (Castelnau et al. 2008). For completeness, a brief description of the theoretical framework of SO-VPSC is presented below. The viscoplastic constitutive behavior is described by the rate-sensitive relation

$$\mathbf{d}(\mathbf{x}) = \sum_s \mathbf{m}^s(\mathbf{x}) \dot{\gamma}^s(\mathbf{x}) = \dot{\gamma}_o \sum_s \mathbf{m}^s(\mathbf{x}) \left| \frac{\mathbf{m}^s(\mathbf{x}) : \mathbf{s}(\mathbf{x})}{\tau_c^s(\mathbf{x})} \right|^n \text{sign}(\mathbf{m}^s(\mathbf{x}) : \mathbf{s}(\mathbf{x})), \tag{19}$$

where $2\mathbf{m}^s = \mathbf{n}^s \otimes \mathbf{b}^s + \mathbf{b}^s \otimes \mathbf{n}^s$ is defined as a symmetric tensor, with $\mathbf{n}^s$ and $\mathbf{b}^s$ the normals to the slip systems' glide plane and the Burgers' vector, respectively. Quantity $\dot{\gamma}^s$ represents the strain-rate accommodated by the slip system $s$, and $\dot{\gamma}_o$ is a normalization factor. The stress exponent $n$ is the rate-sensitivity exponent and $\tau_c^s$ is the critical resolved shear stress of the slip system $s$. The tensors $\mathbf{d}$ and $\mathbf{s}$ are the local strain-rate and deviatoric stress, respectively. Linearizing Equation (19), we obtain

$$\mathbf{d}(\mathbf{x}) = \mathbf{l} : \mathbf{s}(\mathbf{x}) + \mathbf{d}^0 \qquad (20a)$$
$$\mathbf{D} = \mathbf{L} : \mathbf{S} + \mathbf{D}^0 \qquad (20b)$$

where $\mathbf{l}$ and $\mathbf{L}$ are the viscoplastic compliances, $\mathbf{d}^0$ and $\mathbf{D}^0$ are the back-extrapolated terms at the level of the grain and of the aggregate, respectively, and $\mathbf{D}$ and $\mathbf{S}$ are the macroscopic deviatoric strain-rate and stress tensors.

The effective stress potential of the polycrystal described by Equation (20b) may be written in the form

$$U_T = \frac{1}{2}\mathbf{L} :: (\mathbf{S} \otimes \mathbf{S}) + \mathbf{D}^0 : \mathbf{S} + \frac{1}{2}G \qquad (21)$$

which expresses the effective potential $U_T$ of the nonlinear viscoplastic polycrystal in terms of a linearly viscous aggregate with properties determined from variational principles. The last term $G/2$ in Equation (21) is the energy under zero applied stress. The average second-order moment of the stress field is the fourth-order tensor

$$\langle \mathbf{s} \otimes \mathbf{s} \rangle = \frac{2}{c}\frac{\partial U_T}{\partial \mathbf{l}} = \frac{1}{c}\frac{\partial \mathbf{L}}{\partial \mathbf{l}} :: (\mathbf{S} \otimes \mathbf{S}) + \frac{1}{c}\frac{\partial \mathbf{D}^0}{\partial \mathbf{l}} : \mathbf{S} + \frac{1}{c}\frac{\partial G}{\partial \mathbf{l}} \qquad (22)$$

where $c$ is the volume fraction of a given grain. From the average second-order moments of the stress, the associated second-order of the strain-rate can then be evaluated as

$$\langle \mathbf{d} \otimes \mathbf{d} \rangle = (\mathbf{l} \otimes \mathbf{l}) :: \langle \mathbf{s} \otimes \mathbf{s} \rangle + \mathbf{d} \otimes \mathbf{d}^0 + \mathbf{d}^0 \otimes \mathbf{d} - \mathbf{d}^0 \otimes \mathbf{d}^0. \qquad (23)$$

The average second-order moments of the stress field over each grain are obtained by calculating the derivatives in Equation (22). The implementation of the second-order procedure in VPSC follows the work of Liu and Ponte Castañeda (2004). For more details, see the VPSC7c manual (Tomé and Lebensohn 2012).

### 2.3.1 Polycrystal Equipotential Surface

The anisotropy of a viscoplastic material can be described by comparing points that belong to the same reference equipotential surface. This requires probing the material in a given stress direction, while ensuring that the associated dissipation rate will be the same regardless of the chosen direction. This reference Polycrystal Equipotential Surface (PES) is the locus of all stress states associated with a polycrystal with a given texture. In the present study, we use the SO-VPSC polycrystal plasticity model to estimate the PES.

The plastic potential is essentially a function of the stress that can be differentiated to derive the plastic strain-rate. This function $f(\mathbf{S})$ is defined by a constant plastic work rate, $\dot{W}_0$, along the potential. The plastic work rate for an arbitrary macroscopic strain-rate, $\mathbf{D}_0$, is defined as

$$f(\mathbf{S}) = \mathbf{S} : \mathbf{D} = \mathbf{S}_0 : \mathbf{D}_0 = \dot{W}_0 \qquad (24)$$

where $\mathbf{S}_0$ is the stress state corresponding to a given $\mathbf{D}_0$ (or inversely). This function defines a series of convex surfaces in the deviatoric stress space, which are equipotential surfaces when $f(\mathbf{S})$ is constant. As $\mathbf{S}$ or $\mathbf{D}$ is not known *a priori*, the obtained plastic potential rate can be different from $\dot{W}_0$. In this case, the stress point does not lie on the selected equipotential. The stress or strain-rate on the selected equipotential can be obtained based on Hutchinson (1976), which showed that when the magnitude of the strain-rate is changed by a factor $\zeta$, the stress response of the polycrystal becomes

$$\mathbf{S}(\zeta \mathbf{D}) = \zeta^n \mathbf{S}(\mathbf{D}). \qquad (25)$$

As a consequence, the magnitude of $\mathbf{S}$ or $\mathbf{D}$ can be scaled as follows: $\mathbf{D}^* = \mathbf{D}(\dot{W}_0/\dot{W})^{n/(1+n)}$ and $\mathbf{S}^* = \mathbf{S}(\dot{W}_0/\dot{W})^{1/(1+n)}$. The standard VPSC code imposes the strain-rate vectors $\mathbf{D}$ and calculates the associated stress **S** for probing the material response. As mentioned above, both tensors can be renormalized to give the same dissipation rate for every point of the yield surface. The test direction in the strain-rate space is given by $n_{ij} = D_{ij}/\|\mathbf{D}\|$, where $\|\mathbf{D}\| = \sqrt{D_{ij}D_{ij}}$ defines the length of the strain-rate tensor **D**. Thus, the strain-rate tensor can be characterized by a polar representation, which consists of a radius in strain-rate (deviatoric stress) space, $\|\mathbf{D}\|$, and a set of direction cosines ($n_{ij}$). In this representation $\|\mathbf{D}\|$ is an independent variable, whereas the nine values of $n_{ij}$ are not. Since **D** is a symmetric deviatoric second-order tensor, it is possible to represent the strain-rate tensor by a five-component unit vector using a second-order tensor with an orthonormal symmetric base $n_{ij} = n_\lambda b_{ij}^\lambda$, $\lambda = 1, \ldots, 5$ (see Appendix B). Using generalized spherical coordinates, the vector **n** can be expressed as

$$n_{(1)} = \cos\theta_1 \sin\theta_2 \sin\theta_3 \sin\theta_4 \sin\theta_5, \quad n_{(2)} = \cos\theta_2 \sin\theta_3 \sin\theta_4 \sin\theta_5,$$
$$n_{(3)} = \cos\theta_3 \sin\theta_4 \sin\theta_5, \quad n_{(4)} = \cos\theta_4 \sin\theta_5, \quad (26)$$
$$n_{(5)} = \cos\theta_5$$

with $-\pi \leq \theta \leq \pi$ or $0 \leq \theta \leq \pi/2$ for centro or non-centro symmetric evaluation, respectively. As an example, if simulations are restricted to the $S_1 - S_2$ subspace, thus, $\theta_1 = 0$, $-\pi \leq \theta_2 \leq \pi$, $\theta_3 = \theta_4 = \theta_5 = \pi/2$, and Equation (26) reduces to $n_{(1)} = \sin\theta_2$, $n_{(2)} = \cos\theta_2$ $n_{(3)} = 0$, $n_{(4)} = 0$, and $n_{(5)} = 0$ with $\theta_2$ scanned in steps of 15° and the probed strain-rate (deviatoric stress) tensor set to $\|\mathbf{D}\| = 1$. Here, we extend the original implementation of the PES calculation in the VPSC code (Tomé and Lebensohn (2012) – VPSC7c manual) by allowing to choose whether the sampling points will be equispaced in strain-rate or in stress.

As an example, we present in Figure 1 the $\{\pi\}$- and shear projections of the PES for an olivine aggregate with a typical orthorhombic texture, described by 1000 orientations. The orientation distribution function (ODF) and texture intensity (J-index) were quantified using the open-source MTEX toolbox (Mainprice et al. 2015). The olivine slip systems data in Table 1 are derived from experiments at high temperature and moderate pressure conditions (Bai et al. 1991).

**Table 1** Slip systems parameters used in the SO-VPSC simulations. [a]Adimensional values; normalized by flow stress of the (010)[100] slip system; [b]Slip systems not active in olivine, used for stabilizing the calculations, but accommodating ≪ 5 % strain in all simulations.

| Slip Systems | Critical Resolved Shear Stress[a] | Stress Exponent |
|---|---|---|
| (010)[100] | 1 | 3 |
| (001)[100] | 1 | 3 |
| (010)[001] | 2 | 3 |
| (100)[001] | 3 | 3 |
| (011)[100] | 4 | 3 |
| (110)[001] | 6 | 3 |
| {111}⟨110⟩[b] | 50 | 3 |
| {111}⟨011⟩[b] | 50 | 3 |
| {111}⟨101⟩[b] | 50 | 3 |

All curves are normalized by the work rate of an isotropic olivine polycrystal deformed under similar conditions (full symbols). No appreciable differences are obtained using either strain-rate or stress sampling for the calculation of the yield surfaces. Since only intracrystalline glide deformation modes are taken into account (no twinning), the predicted surfaces are centro-symmetric. The four equipotential surfaces are then fitted using a least-square method to obtain the six coefficients ($F$, $G$, $H$, $L$, $M$, $N$) that satisfy the anisotropic Hill yield function (see Equation (9)). Table 2 presents the fitted coefficients for the orthorhombic texture shown in Figure 1.

| Sampling variable | $F$ | $G$ | $H$ | $L$ | $M$ | $N$ |
|---|---|---|---|---|---|---|
| Strain-rate (err = 0.034) | 0.0220 | 0.2208 | 0.3769 | 8.6133 | 2.0777 | 2.3101 |
| Stress (err = 0.028) | 0.0225 | 0.2275 | 0.3744 | 8.9183 | 2.1258 | 2.3016 |

**Table 2** Hill coefficients obtained by imposing either strain-rate or stress for sampling the PES

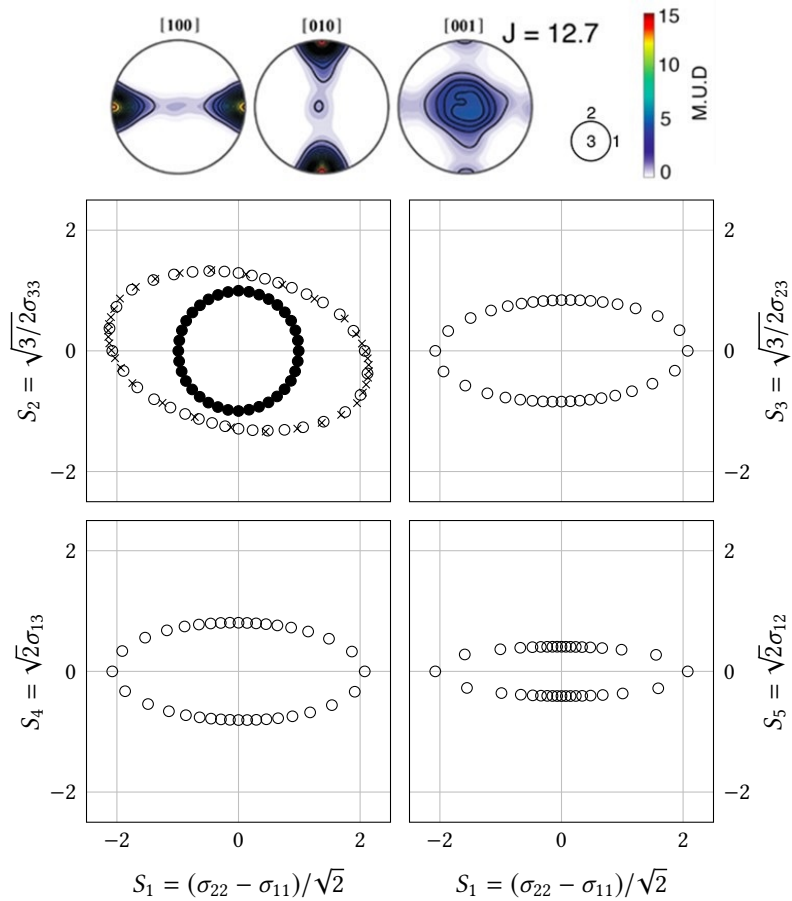

**Figure 1**  {π}- and shear projections of the equipotential yield surface for a textured olivine polycrystal calculated using the SO-VPSC approach. Simulations were run using slip systems data compatible with deformation at high temperature and moderate pressure conditions, see Table 1. The yield surface was sampled using 36 equispaced ($\Delta\theta = 15°$) stress points (open circles). For comparison, the yield surface obtained using a strain-rate sampling is also displayed in the {π}-projection (crosses). All loading conditions are normalized by the work rate displayed by an isotropic olivine polycrystal under similar conditions–the corresponding equipotential is displayed in the {π}-projection as full circles. The insert on the top shows the olivine CPO used in the anisotropic calculations. The pole figure (contours in multiples of a uniform distribution) and the texture intensity (J-index) were processed using the open-source MTEX toolbox (Mainprice et al. 2015).

### 2.3.2 Model validation

In this section, we compare Adeli3D-anis predictions with those obtained directly by SO-VPSC simulations for two simple case studies: simple shear or axial compression applied to a cube with a strong, but homogeneous olivine texture at various orientations relative to the mechanical solicitation. The slip systems data and the olivine texture used in both simulations are that reported in Table 1 and Figure 1. Both models employ a large deformation formalism. The Hill constitutive material law is integrated in the material texture reference frame $\hat{\mathcal{R}}$ (i.e., Hill coefficients are defined in $\hat{\mathcal{R}}$).

Since texture and hence viscous anisotropy evolution as a function of strain are not implemented in the current version of the finite-element simulations, Von Mises equivalent (VM) stresses $s_{eq} = \sqrt{3s_{ij}s_{ij}/2}$ predicted by the Adeli3D-anis at steady-state are compared to VM stresses predicted by the SO-VPSC at the end of a single deformation step, before reorientation of the texture. The VM stress in Adeli3D-anis is averaged over all (six) elements of the cube mesh. The orientation of the texture reference frame (XYZ) relative to the mechanical solicitation reference frame (123) is varied from 0° to 90° at 15° intervals. SO-VPSC simulations are adimensionalized: all stresses are normalized by the stress in the easiest (010)[100] slip system of olivine, whereas Adeli3D-anis produces absolute stresses that depend on the isotropic rheological parameters and on the imposed temperature. To compare the results of Adeli3D-anis and SO-VPSC models, the VM stresses for the textured polycrystal were therefore normalized by the VM of an olivine polycrystal with 1000 randomly oriented grains, which has an isotropic mechanical response.

Two sets of boundary conditions were considered in the axial extension tests. For the first

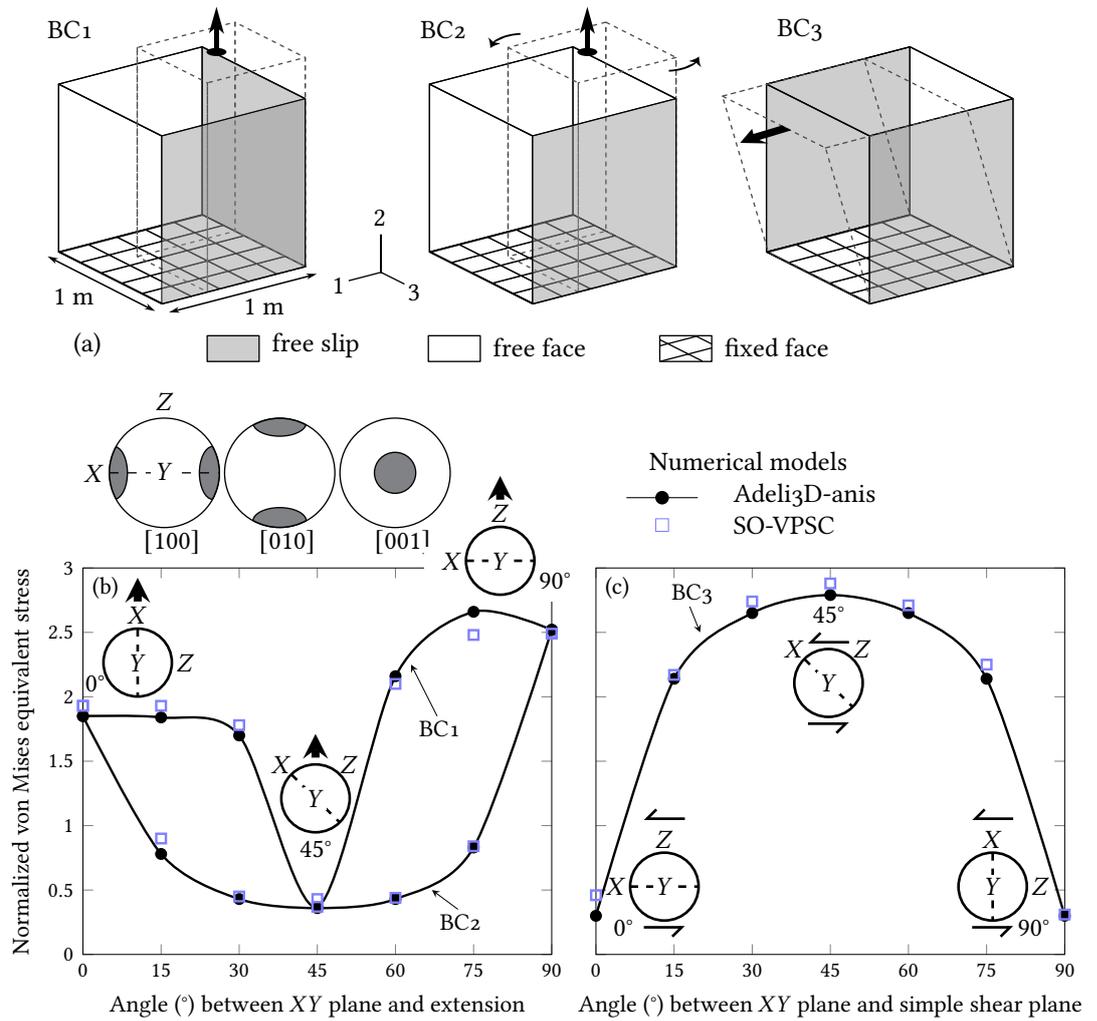

**Figure 2** (a) Imposed boundary conditions in Adeli3D-anis simulations and velocity gradient tensors (**L**) imposed in the corresponding SO-VPSC simulations: $\mathbf{L}_{ext}$ extension and $\mathbf{L}_{ss}$ simple shear. BC: boundary conditions. Comparison of the predictions of anisotropic ADELI3D and SO-VPSC simulations: (b) axial extensional tests with two sets of boundary conditions (BC1 = two planes of symmetry and BC2 = a single plane of symmetry); (c) simple shear test (BC3). Parameters describing the isotropic and anisotropic part of the rheology in the Adeli3D-anis simulations are presented in Table 3 and Table 2 (stress sampling), respectively.

set (BC1 in Figure 2(a)), extension was imposed in the Adeli3D-anis simulation by applying a constant velocity normal to the face normal to axis 2 of the cube and keeping the opposite face fixed, one of the faces normal to axes 1 and 3 is a symmetry plane (free slip conditions) and the other is free (free face). This correspond to mixed boundary conditions in SO-VPSC simulations, where an extensional velocity 22 is imposed, all shear velocity components are imposed null ($L_{i \neq j} = 0$), and equal non-null stresses 11 and 33 are imposed. The corresponding tensor reads:

$$\mathbf{L}_{1-ext} = \begin{bmatrix} * & 0 & 0 \\ 0 & 0.8 & 0 \\ 0 & 0 & * \end{bmatrix} \quad (27)$$

where the symbol * indicates that the magnitude of the component is unknown and must be determined as a computational result. In the second set of boundary conditions (BC2 in Figure 2(a)), the two faces normal to axis 1 are free in the Adeli3D-anis simulations. This corresponds to SO-VPSC simulations where the extensional velocity 22, a null velocity to half of the shear components ($L_{12} = L_{13} = L_{23} = 0$), equal non null stresses 11 and 33, and null shear stresses. The predicted solutions for all loading-geometries are remarkably similar between the Adeli3D-anis and the SO-VPSC models for the two sets of boundary conditions, Figure 2(b). The corresponding tensor reads:

$$\mathbf{L}_{2-\text{ext}} = \begin{bmatrix} * & * & * \\ 0 & 0.8 & * \\ 0 & 0 & * \end{bmatrix}. \tag{28}$$

The less stringent boundary conditions (BC2), i.e. allowing a rigid rotation, results in lower normalized VM stresses for all solicitations oblique to the texture reference frame, except at 45°, see Figure 2(b). In our case, due to the symmetry of the initial olivine texture, this rigid rotation may only occur around the extension axis.

Simple shear tests were performed by applying a constant tangential velocity parallel to axis 1 on the face normal to axis 2, keeping the opposite face fixed, and imposing null normal velocities to the two faces normal to axis 3 (BC3 in Figure 2(a)). The corresponding tensor reads:

$$\mathbf{L}_{\text{SS}} = \begin{bmatrix} 0 & 1.4 & 0 \\ 0 & 0 & 0 \\ 0 & 0 & 0 \end{bmatrix}. \tag{29}$$

Equivalent boundary conditions are simulated in SO-VPSC by imposing a non-null component 12 and null values to all other components of the velocity gradient tensor. The stress variation as a function of the orientation of the imposed shear relative to the texture reference frame predicted by Adeli3D-anis and SO-VPSC models are also remarkably similar (Figure 2(c)), confirming that the present parameterization based on the Hill (1948) yield function is suitable for describing the viscous anisotropy of an olivine polycrystal with an orthotropic texture.

| $E$ (GPa) | $\nu$ | $n$ | $\gamma_0$ (Pa$^{-n}$ s$^{-1}$) | $Q$ (kJ/mol) | $T$ (K) |
|---|---|---|---|---|---|
| 100 | 0.25 | 3 | $10^{-18}$ | 500 | 1423 |

**Table 3** Parameters describing the isotropic part of the rheology in the validation tests.

## 3 Application: effect of texture-induced viscous anisotropy associated with fossil shear zones on the deformation of a continental plate

Viscoplastic deformation of mantle rocks in lithospheric-scale shear zones, i.e., narrow zones accommodating shear displacements between relatively undeformed domains of a tectonic plate, leads to development of olivine textures that may be preserved for very long time spans (hundreds of millions years, *cf.* (Tommasi and Vauchez 2015)). Anisotropic viscosity due to fossil olivine texture in mantle shear zones has been argued to trigger localized deformation in the plates when the mechanical solicitation is oblique to the trend of the shear zones, leading to the formation of new plate boundaries parallel to these ancient structures (Vauchez et al. 1997; Tommasi and Vauchez 2001; Tommasi et al. 2009). However, previous simulations testing these effects, which directly coupled VPSC polycrystal plasticity models into the finite-element codes simulating the geodynamical flows, were too computationally demanding for full investigation of the interactions between texture-induced anisotropy and other strain localization processes active on Earth. The parameterization presented here allows for a significant gain in both computation time and memory requirements, enabling to run 3D geodynamical models that explicitly consider the effect of texture-induced viscous anisotropy in the mantle on the plates' dynamics. Its first application in geodynamics (Mameri et al. 2020) focused on investigating the possible role of texture-induced viscous anisotropy in the mantle in producing enigmatic alignments of active seismicity in intraplate settings. This work testifies that texture-induced viscous anisotropy in the mantle may produce strain localization not only in the mantle, but also in the overlying crust (Figure 3). A detailed analysis of the geological aspects of the problem are presented in (Mameri et al. 2020), but these simulations allowed to investigate the feedbacks between localized deformation controlled by texture-induced viscous anisotropy in the mantle and the deformation processes in the brittle (plastic) upper crust. Parameters controlling the isotropic part of the mantle and crust rheologies are shown in Table 4.

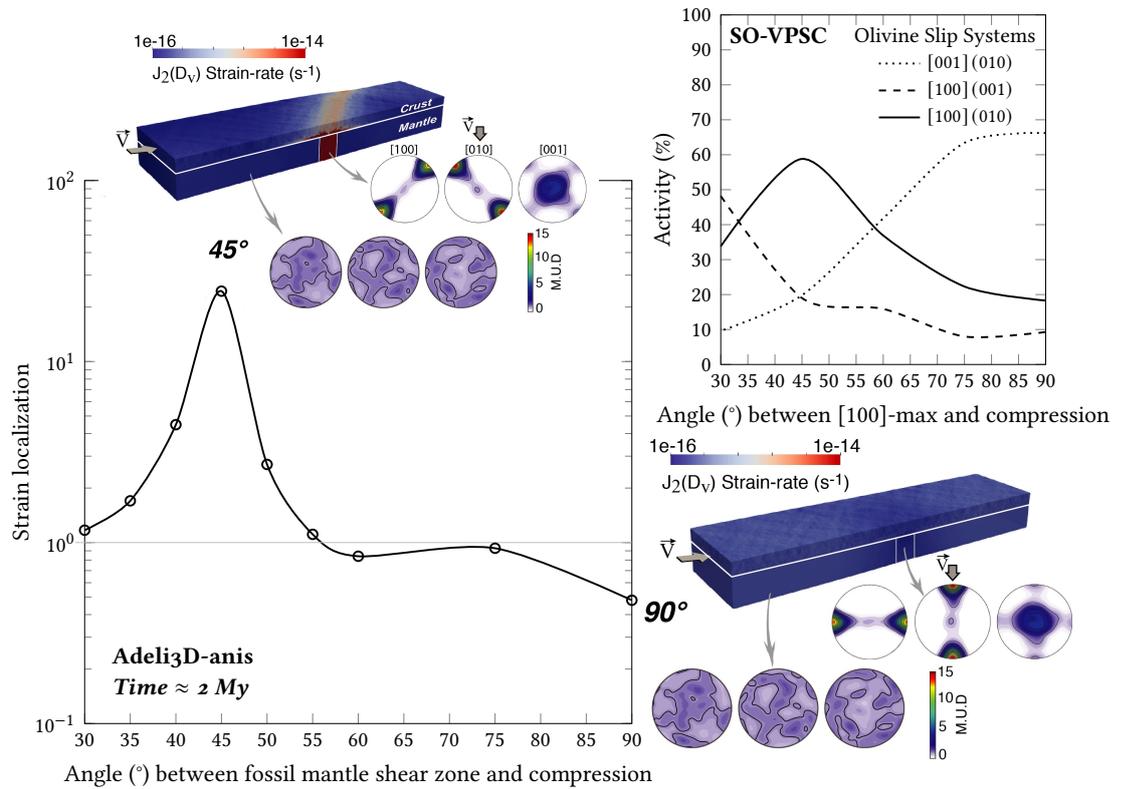

**Figure 3**   Strain localization quantified as the ratio between the average of second-invariant of the strain-rate within the fossil mantle shear zone and outside it for different orientations of the fossil shear zone relatively to the imposed shortening. The model represents a 1100 km long, 500 km wide, and 120 km thick continental plate containing a fossil shear zone marked by a change in viscous anisotropy simulating a change with the olivine texture in the lithospheric mantle. The Hill parameters in the fossil shear zone are coherent an olivine texture formed in response to a past strike-slip deformation (horizontal shear in a vertical plane). The surrounding mantle has Hill parameters simulating a random texture. Strain localizes in the fossil shear zone when it is oblique to the imposed compression, but strain localization is not symmetrical with respect to the orientation of the fossil shear zone relative to the imposed compression. These predictions reflect correctly the contrast in CRSS between the different slip systems in olivine, see Table 1, and their relative activity as predicted by SO-VPSC simulations considering a polycrystal with the texture (orientation and intensity) used to define the viscous anisotropy in the fossil shear zone for each model (insert). Maximum strain localization occurs when the fossil shear zone is at 45° to the compression, where shear stresses resolved onto the easy [100](010) slip system in most olivine crystals are maximum. The lower strain localization in simulations with the fossil shear zone at 60° to the compression relative to that in simulations with the fossil shear zone at 30° to the compression is consistent with the higher activity of the hard [001](010) slip system in the former.

## 4 Computational cost

Figure 4 compares the cost associated with using the present anisotropic Maxwell rheology relatively to the classical isotropic formulation for simulations with increasing number of elements. Plane strain compression is imposed to a plate with either a homogeneous isotropic rheology (Adeli3D) or Hill parameters simulating a random texture (Adeli3D-anis). Parameters controlling the isotropic part of the rheology are the same in the two simulations (Table 4). Two different tolerance values for convergence in the flow law integration were tested. Both isotropic and anisotropic computation times increase almost linearly with the number of elements. Thus, the ratio between the CPU time for the isotropic and the anisotropic simulations does not depend significantly on the mesh size. It even decreases with increasing mesh size, stabilizing for fine meshes. If more severe convergence tolerance criteria are imposed, the cost of the anisotropic rheology is slightly higher. In all cases, the proposed anisotropic parameterization induces an additional numerical effort smaller than three times the computational cost of the isotropic rheology. It should be noted that this ratio may change slightly depending on the degree of anisotropy induced by the texture of the material.

**Table 4** Isotropic material parameters used in the geodynamical simulations: *frictional weakening is imposed and $\phi$ is reduced to 15° when plastic strain in the mesh element exceeds 1 %. References: [a]Paterson and Luan (1990), [b]Chopra and Paterson (1984).

| | | Wet quartzite[a] | Wet dunite[b] |
|---|---|---|---|
| Density | $\rho$ (kg m$^{-3}$) | 2653 | 3300 |
| Young modulus | $E$ (GPa) | 70 | 160 |
| Poisson ratio | $\nu$ | 0.25 | 0.28 |
| Fluidity | $\gamma_0$ (Pa$^{-n}$ s$^1$) | $1.63 \times 10^{-26}$ | $3.98 \times 10^{-25}$ |
| Activation Energy | $Q$ (kJ mol$^{-1}$ K$^{-1}$) | 135 | 498 |
| Stress exponent | $n$ | 3.1 | 4.5 |
| Angle of friction* | $\phi$ | 30 | – |
| Cohesion | $c$ (MPa) | 10 | – |

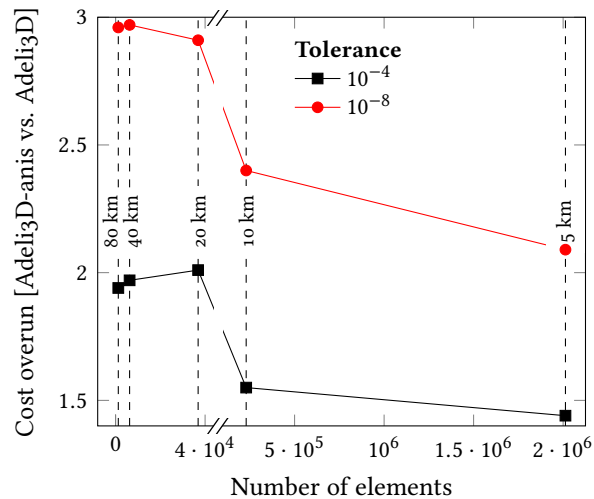

**Figure 4** Evolution of ratio in the CPU time between simulations using the present anisotropic rheology (Adeli3D-anis) and a classical isotropic formulation (Adeli3D) for increasingly finer mesh sizes, from 80 km to 5 km. The model domain is 1100 km long, 550 km wide, and 120 km thick. Adeli3D-anis simulations were performed using Hill coefficients for an isotropic texture ($F = G = H = 1/2$, $L = M = N = 3/2$).

## 5 Conclusion

We developed a relatively simple formulation for a Maxwell rheology combining an isotropic elastic and a texture-induced anisotropic non-linear viscous behavior, parameterized based on the Hill (1948) orthotropic yield criterion. The six Hill yield surface coefficients are obtained by least-square fitting of four equipotential surfaces calculated using the SO-VPSC model. This formulation was implemented in the 3D thermo-mechanical code Adeli3D developed for modeling geodynamical flows. The numerical integration technique associated with the anisotropic viscous parameterization was validated by recovering the semi-analytical solution for a shear test either assuming a linear ($n = 1$) or a non-linear ($n = 3.6$) viscous rheology. Comparison of the predictions of Adeli3D-anis for simple shear and axial extension of a cube with a homogeneous olivine texture further validated the implementation. The computational effort only increases by a factor of 2-3 with respect to an equivalent simulation using a isotropic Maxwell rheology. An example of a geodynamical application of this parameterized viscous anisotropy is presented. These models allow to quantifying the effect of texture-induced viscous anisotropy in the mantle on the dynamics of tectonic plates. They predict coupling between localized deformation due to lateral variations in the orientation and intensity of olivine texture in the mantle and the strain distribution in the shallow crust. A current limitation for using the present parameterization for modeling more complex geodynamical flows is our ability to simulate the evolution of the anisotropy due to the evolution of the texture in response to viscous deformation while retaining the computational efficiency. This is part of an ongoing study, where different strategies are evaluated.

## A Validation of the implemented Hill-based parameterization of the viscoelastic anisotropy

Before applying it to characterize the texture-induced viscous anisotropy of olivine polycrystals, we validated the implementation by comparison with semi-analytical solutions for simple settings.

The numerical integration technique associated with the anisotropic viscous parameterization presented in the previous section was first validated based on its ability to recover a semi-analytical solution for a shear test applied to a material with either a linear ($n = 1$) or a non-linear ($n = 3.6$) viscous rheology. Shear is imposed by a velocity field described in the laboratory reference frame $\mathcal{R}$ (axes $x, y, z$) by $\mathbf{v}_{[\mathcal{R}]}(x, y, z) = (2ay, 0, 0)$, with $a > 0$. The material reference frame $\hat{\mathcal{R}}$ (anisotropy axes $x_1, x_2, x_3 = z$) is rotated from the laboratory reference frame $\mathcal{R}$ by an angle $\theta$ (Figure A.1(a)). The strain-rate, the co-rotational rotation rate and the deviatoric stress are expressed in $\hat{\mathcal{R}}$ as follows:

$$\hat{\mathbf{D}} = a \begin{bmatrix} \sin 2\theta & \cos 2\theta & 0 \\ \cos 2\theta & -\sin 2\theta & 0 \\ 0 & 0 & 0 \end{bmatrix}, \quad \hat{\omega} = a \begin{bmatrix} 0 & 1 & 0 \\ -1 & 0 & 0 \\ 0 & 0 & 0 \end{bmatrix}, \quad \hat{\mathbf{S}} = \begin{bmatrix} \hat{s}_{11} & \hat{s}_{12} & 0 \\ \hat{s}_{12} & \hat{s}_{22} & 0 \\ 0 & 0 & -(\hat{s}_{11} + \hat{s}_{22}) \end{bmatrix} \quad (A.1)$$

and in $\mathcal{R}$ the stress components are given by

$$\mathbf{S} = \mathbf{R}_0 \hat{\mathbf{S}} \mathbf{R}_0^\top = \begin{bmatrix} \sigma_{xx} & \sigma_{xy} & 0 \\ \sigma_{xy} & \sigma_{yy} & 0 \\ 0 & 0 & -(\sigma_{xx} + \sigma_{yy}) \end{bmatrix} \quad (A.2)$$

with $\mathbf{R}_0$, the rotation matrix from $\hat{\mathcal{R}}$ to $\mathcal{R}$. In this simple case, from Equation (12) and assuming zero initial stresses, the resulting differential system can be written as

$$\begin{cases} \dot{\mathbf{y}}(t) + \mathbf{C}_n(\mathbf{y})\mathbf{y}(t) = \mathbf{b}, & 0 < t \leq t_f \\ \mathbf{y}(0) = \mathbf{0} \end{cases} \quad (A.3)$$

where $\mathbf{y} = (\hat{s}_{11}, \hat{s}_{22}, \hat{s}_{12})^\top$, $\mathbf{b} = 2\mu a (\sin 2\theta, -\sin 2\theta, \cos 2\theta)^\top$, and

$$\mathbf{C}_n(\mathbf{y}) = \begin{bmatrix} (F + 2H)\alpha_n(\mathbf{y}) & (H - F)\alpha_n(\mathbf{y}) & -2a \\ (G - F)\alpha_n(\mathbf{y}) & (F + 2G)\alpha_n(\mathbf{y}) & 2a \\ a & -a & L\alpha_n(\mathbf{y}) \end{bmatrix} \quad (A.4)$$

with $\alpha_n(\mathbf{y}) = \frac{4}{3}\mu\gamma(F(y_1 - y_2)^2 + G(y_1 + 2y_2)^2 + H(2y_1 + y_2)^2 + 2Ly_3^2)^{(n-1)/2}$.

We can make the following remarks:

1. For the linear case ($n = 1$) the coefficients of $\mathbf{C}_n$ are obviously constant and the solution of the system takes the form $\mathbf{y} = (\mathbf{I} - e^{-t\mathbf{C}_1})\mathbf{C}_1^{-1}\mathbf{b}$ while for the general non-linear case, only a numerical solution can be computed (for these validations tests, we used an ODE solver of Matlab).
2. The components 13, 23, 31 and 32 of $\mathbf{C}_n$ (implying the magnitude of the velocity gradient $a$) come from the finite strain formalism (Jaumann derivative).
3. Unlike the isotropic case for which $\sigma_{xx}$ and $\sigma_{yy}$ are always opposite, an out-of-plane stress $\sigma_{zz}$ exists whenever $H \neq G$.

Validation tests in Adeli3D-anis were run for a cube of $1\,\text{m}^3$ composed of six four–node tetrahedral finite-elements with homogeneous isotropic and anisotropic material parameters (Figure A.1(b) and Figure A.1(c)). Shear deformation was imposed by applying a constant tangential velocity parallel to axis 1 on the face normal to axis 2 of the cube, while keeping the opposite face fixed and imposing null normal velocities to the two faces normal to axis 3 (Figure 3). The isotropic material parameters are given in Table A.1.

For both isotropic linear and anisotropic non-linear cases, the Adeli3D-anis and the semi-analytical solution are numerically identical (Figure A.1(b) and Figure A.1(c)). In both cases, the imposed simple shear deformation is associated with a dominant $\sigma_{xy}$ stress component, but the presence of a texture-induced anisotropy results in enhancement of this shear component relatively to the diagonal ones. As it is mentioned above anisotropy ($H \neq G$) induces a 3D stress field, where an out-of-plane stress $\sigma_{zz}$ takes place (Figure A.1(c)).

|       | $\lambda$ (GPa) | $\mu$ (GPa) | $n$ | $\dot{\gamma}_0$ (Pa$^{-n}$ s$^{-1}$) | $Q$ (kJ mol$^{-1}$) | $T$ (K) |
|-------|-----------------|-------------|-----|----------------------------------------|---------------------|---------|
| Test 1 | 40 | 40 | 1   | $0.5 \times 10^{-12}$ | 500 | 1423 |
| Test 2 | 40 | 40 | 3.6 | $0.5 \times 10^{-18}$ | 500 | 1423 |

**Table A.1** Parameters describing the isotropic part of the rheology in the validation tests.

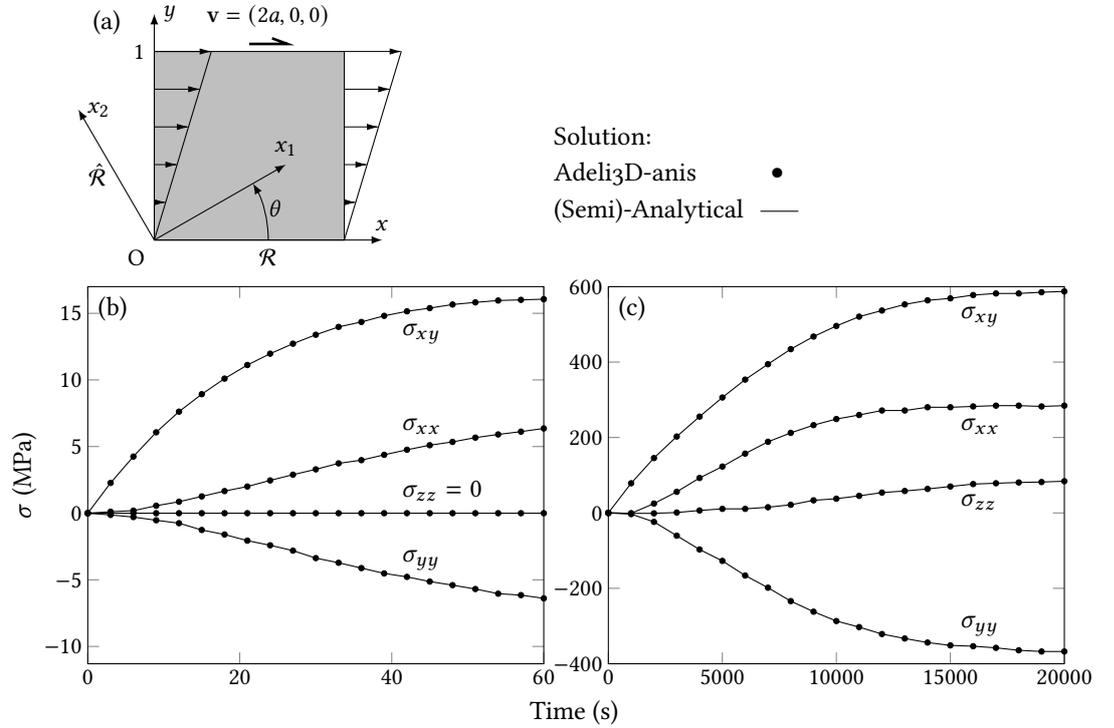

**Figure A.1** (a) Simple shear tests for validating the implementation of the parameterization of the viscous anisotropy in large deformations for the general case of a misalignment $\theta = 30°$ between the anisotropy axes ($\hat{\mathcal{R}}$) with respect to the laboratory axes ($\mathcal{R}$). Components of the stress tensor as a function of time for the semi-analytical solution and the Hill-based parameterization of the viscoplastic anisotropy implemented in Adeli3D-anis: (b) Newtonian isotropic case (von Mises: $F = 0.5$, $L = 1.5$, $G = 0.5$, $M = 1.5$, $H = 0.5$, $N = 1.5$) with $n = 1$ and $a = 10^{-2}$ s$^{-1}$; (c) non-Newtonian anisotropic case (Hill: $F = 0.0$, $L = 8.9$, $G = 0.2$, $M = 2.1$, $H = 0.3$, $N = 2.2$) with $n = 3.6$, $a = 10^{-6}$ s$^{-1}$, and $\theta = 30°$. The Hill yield surface coefficients are defined in the material texture reference frame $\hat{\mathcal{R}}$, which is orthotropic. They correspond to those of an isotropic and a strongly textured olivine polycrystal, determined following the approach described in Section 2.3. Shear is imposed parallel to the maximum concentration of [100] of the texture. The factor $a$ is proportional to the norm of the imposed velocity gradient.

## B Orthonormal base for second- and fourth-order symmetric tensors

When dealing with incompressible media, it is convenient to explicitly decompose stress and strain-rate into deviatoric and hydrostatic components and to confine them to different subspaces, which may be decoupled for certain mechanical regimes. There are different ways of achieving such a decomposition. In the present calculations, we express the second-order tensorial quantities in an orthonormal basis of second-order symmetric tensors $\mathbf{b}^\lambda$, defined as

$$\mathbf{b}^1 = \frac{1}{\sqrt{2}} \begin{bmatrix} -1 & 0 & 0 \\ 0 & 1 & 0 \\ 0 & 0 & 0 \end{bmatrix}, \quad \mathbf{b}^2 = \frac{1}{\sqrt{6}} \begin{bmatrix} -1 & 0 & 0 \\ 0 & -1 & 0 \\ 0 & 0 & 2 \end{bmatrix}, \quad \mathbf{b}^3 = \frac{1}{\sqrt{2}} \begin{bmatrix} 0 & 0 & 0 \\ 0 & 0 & 1 \\ 0 & 1 & 0 \end{bmatrix}, \quad \text{(B.1)}$$

and

$$\mathbf{b}^4 = \frac{1}{\sqrt{2}} \begin{bmatrix} 0 & 0 & 1 \\ 0 & 0 & 0 \\ 1 & 0 & 0 \end{bmatrix}, \quad \mathbf{b}^5 = \frac{1}{\sqrt{2}} \begin{bmatrix} 0 & 1 & 0 \\ 1 & 0 & 0 \\ 0 & 0 & 0 \end{bmatrix}, \quad \mathbf{b}^6 = \frac{1}{\sqrt{3}} \begin{bmatrix} 1 & 0 & 0 \\ 0 & 1 & 0 \\ 0 & 0 & 1 \end{bmatrix}. \tag{B.2}$$

The components of this basis have the property $b_{ij}^\lambda b_{ij}^{\lambda'} = \delta_{\lambda\lambda'}$ and provide a unique 'vector' and 'matrix' representation of second- and fourth-order symmetric tensors, respectively. In the particular case of the stress tensor $\sigma_{ij} = \sigma_\lambda b_{ij}^\lambda$ where $\sigma_\lambda = \sigma_{ij} b_{ij}^\lambda$. The orthonormality of the basis guarantees that the six-dimensional strain-rate and stress vectors are work conjugates, i.e.

$$d_\lambda \sigma_\lambda = d_{ij} \sigma_{ij}. \tag{B.3}$$

The explicit form of the six components $(\sigma_1, \sigma_2, \sigma_3, \sigma_4, \sigma_5, \sigma_6)$ of $\sigma_\lambda$ is

$$\left( \frac{\sigma_{22} - \sigma_{11}}{\sqrt{2}}, \frac{2\sigma_{33} - \sigma_{11} - \sigma_{22}}{\sqrt{6}}, \sqrt{2}\sigma_{23}, \sqrt{2}\sigma_{13}, \sqrt{2}\sigma_{12}, \frac{\sigma_{11} + \sigma_{22} + \sigma_{33}}{\sqrt{3}} \right) \tag{B.4}$$

and similarly for the components $d_\lambda$. It is clear that in this representation the first five components are deviatoric and the sixth is proportional to the hydrostatic component of the tensor. conversely, to convert back the vector to the second-order tensor

$$\begin{bmatrix} \sigma_{11} & \sigma_{12} & \sigma_{13} \\ & \sigma_{22} & \sigma_{23} \\ \text{sym} & & \sigma_{33} \end{bmatrix} = \begin{bmatrix} -\frac{\sigma_1}{\sqrt{2}} - \frac{\sigma_2}{\sqrt{6}} + \frac{\sigma_6}{\sqrt{3}} & \frac{\sigma_5}{\sqrt{2}} & \frac{\sigma_4}{\sqrt{2}} \\ & \frac{\sigma_1}{\sqrt{2}} - \frac{\sigma_2}{\sqrt{6}} + \frac{\sigma_6}{\sqrt{3}} & \frac{\sigma_3}{\sqrt{2}} \\ \text{sym} & & \frac{2\sigma_2}{\sqrt{6}} + \frac{\sigma_6}{\sqrt{3}} \end{bmatrix}. \tag{B.5}$$

**Authors' contributions**    J.S.: Conceptualization, Formal Analysis, Model implementation, Resources, Supervision, Funding acquisition. R.H.: Conceptualization, Formal Analysis, Model implementation, Supervision. A.T.: Conceptualization, Formal Analysis, Software, Resources, Supervision, Funding acquisition. L.M.: Formal Analysis, Writing – original draft, Software, Investigation, Simulation and Visualization. All authors: Writing – review and editing and gave final approval of the manuscript.

**Acknowledgements**    We are grateful to Ricardo Lebensohn and Carlos Tomé for making the VPSC7c code freely available. This project received funding from the European Union's Horizon 2020 research and innovation program under the Marie Sklodowska-Curie grant agreement No. 642029 (ITN-CREEP), and from the CNRS (France) and CONICET (Argentina) under the Projet International de Cooperation Scientifique MicroTex (No. 067785). The authors thank the reviewers and editorial board.

**Ethics approval and consent to participate**    Not applicable.

**Consent for publication**    Not applicable.

**Competing interests**    The authors declare that they have no competing interests.

**Journal's Note**    JTCAM remains neutral with regard to jurisdictional claims in published maps and institutional affiliations.